\documentclass[aps,prl,reprint,twocolumn,nofootinbib]{revtex4}

\usepackage{graphicx}

\usepackage{color}\usepackage{enumerate}
\usepackage{amsmath}\usepackage{amssymb}\usepackage{stmaryrd}
\usepackage{multirow}
\usepackage{float}
\usepackage{longtable,booktabs}
\usepackage{color}\usepackage{enumerate}
\usepackage{amsmath}\usepackage{amssymb}\usepackage{stmaryrd}
\usepackage{multirow}
\usepackage{float}
\usepackage[toc,page]{appendix}
\usepackage{longtable,booktabs}
 \usepackage{epsfig}
\usepackage{graphicx} 
\usepackage{soul}

\usepackage{psfrag} 
\usepackage{slashed}
\usepackage{amsmath,amssymb} 
\usepackage{colordvi} 
\usepackage{hyperref}
\usepackage{setspace}
\usepackage[all]{hypcap}
\usepackage{natbib}
\usepackage{subfigure}

\hypersetup{
    bookmarks=true,         
    unicode=false,          
    pdftoolbar=true,        
    pdfmenubar=true,        
    pdffitwindow=false,     
    pdfstartview={FitH},    
    pdftitle={My title},    
    pdfauthor={Author},     
    pdfsubject={Subject},   
    pdfcreator={Creator},   
    pdfproducer={Producer}, 
    pdfkeywords={keyword1} {key2} {key3}, 
    pdfnewwindow=true,      
    colorlinks=true,       
    linkcolor=red,          
    citecolor=green,        
    filecolor=magenta,      
    urlcolor=black           
}
\def\be{\begin{equation}}
\def\bea{\begin{eqnarray}}
\def\eea{\end{eqnarray}}
\def\ee{\end{equation}}

\def\d{\mathrm{d}}

\DeclareMathOperator{\sgn}{sgn}
\DeclareMathOperator{\Tr}{Tr}

\newcommand{\beq}{\begin{equation}}
\newcommand{\eneq}{\end{equation}}

\newcommand{\ket}[1]{\left|#1\right\rangle}

\def\be{\begin{equation}}
\def\ee{\end{equation}}
\def\ba{\begin{eqnarray}}
\def\ea{\end{eqnarray}}

\def\d{\delta}

\def\L{\Lambda}

\input{epsf}
\def\be{\begin{equation}}
\def\bea{\begin{eqnarray}}
\def\eea{\end{eqnarray}}
\def\ee{\end{equation}}

\def\d{\mathrm{d}}

\begin{document}

\title{Electromagnetic Response of Three-dimensional Topological Crystalline Insulators}
\author{Srinidhi T. Ramamurthy}
\author{Yuxuan Wang}
\author{Taylor L. Hughes}
\affiliation{Department of Physics, Institute for Condensed Matter Theory, University of Illinois at Urbana-Champaign, IL 61801, USA}

\begin{abstract}
Topological crystalline insulators (TCI) are a new class of materials which have metallic surface states on select surfaces due to point group crystalline symmetries. In this letter, we consider a model for a three-dimensional (3D) topological crystalline insulator with Dirac nodes occurring on a surface that are protected by the mirror and time reversal symmetry. We demonstrate that the electromagnetic response for such a system is characterized by a $1$-form $b_{\mu}$. $b_{\mu}$ can be inferred from the locations of the surface Dirac nodes in energy-momentum space and couples to the surface Dirac nodes like a valley gauge field. From both the effective action and analytical band structure calculations, we show that the vortex core of $\vec b$ or a domain wall of a component of $\vec b$ can trap surface charges. 
\end{abstract}

\maketitle

Topological phases of matter have been at the forefront of condensed matter physics for the past decade. One reason for the excitement is that topological phases can exhibit electromagnetic responses that display their topological nature. The integer quantum Hall (IQH) effect was the first such system,  and its quantized Hall conductance is characterized by a topological integer \cite{tknn} multiplying the conductance quantum $e^2/h.$ In recent years, the ten-fold, periodic table classification of electronic topological insulators and superconductors with time-reversal (TR) $\mathcal{T}$, particle-hole (PH) $\mathcal{C},$ and/or chiral symmetry $\mathcal{S}$ was completed in Refs.\ \cite{schnyder,qhz2008,kitaev}, and ushered in the concept of a symmetry protected topological (SPT) phase. 
The electromagnetic (EM) response theories of many of the topological insulator (TI) phases were developed in Ref.~\cite{qhz2008}, and extended what was known about the IQH to all fermionic SPTs. For example, the 3D $\mathcal{T}$-invariant topological insulator has an odd number of Dirac cones on each surface, and harbors a half quantum Hall effect when $\mathcal{T}$ is broken on the surface. An odd number of Dirac cones, and the corresponding Hall effect, can never occur in a purely 2D system with the same symmetries without interactions. Indeed, the surface quantum Hall effect is actually a signature of a \emph{bulk} EM response: the topological magneto-electric effect\cite{qhz2008,essin2009} with a response coefficient determined by a $\mathbb{Z}_{2}$ topological invariant\cite{qhz2008}. 

After the periodic table was complete, and after many exciting materials predictions and discoveries\cite{Kane2005A,bernevig2006c,konig2008,hsieh08,moore2009topological,xia2009observation,zhang2009experimental,chang2013}, the classification of topological crystalline phases (TCIs) with point/space-group symmetries, such as reflection and discrete rotation, was initiated and continues to be an active area of research\cite{FuKaneWeak2007,teo2008,futci,hughes2011inversion,turner2012,fang2012bulk,teohughes1,slager2012,Benalcazar2013,Morimoto2013,RyuReflection2013,fang2013entanglement,HughesYaoQi13,Sato2013,Kane2013Mirror,jadaun2013,slager2013}. One highlight of this line of research was the prediction and experimental confirmation of a 3D TCI phase in PbSnTe\cite{hsiehtci,tanakatci,xu2012,dziawatci}. The topological properties of this system are protected by mirror symmetry, and it exhibits an insulating bulk with an even number of symmetry-protected Dirac-cone surface states on mirror-symmetric surfaces. The goal of this article is to predict a characteristic electromagnetic response property that can be observed in PbSnTe and similar 3D TCIs protected by mirror symmetry (mTCIs).

\begin{figure}[!h]
\label{fig:tciillustration}
  \includegraphics[width=6cm]{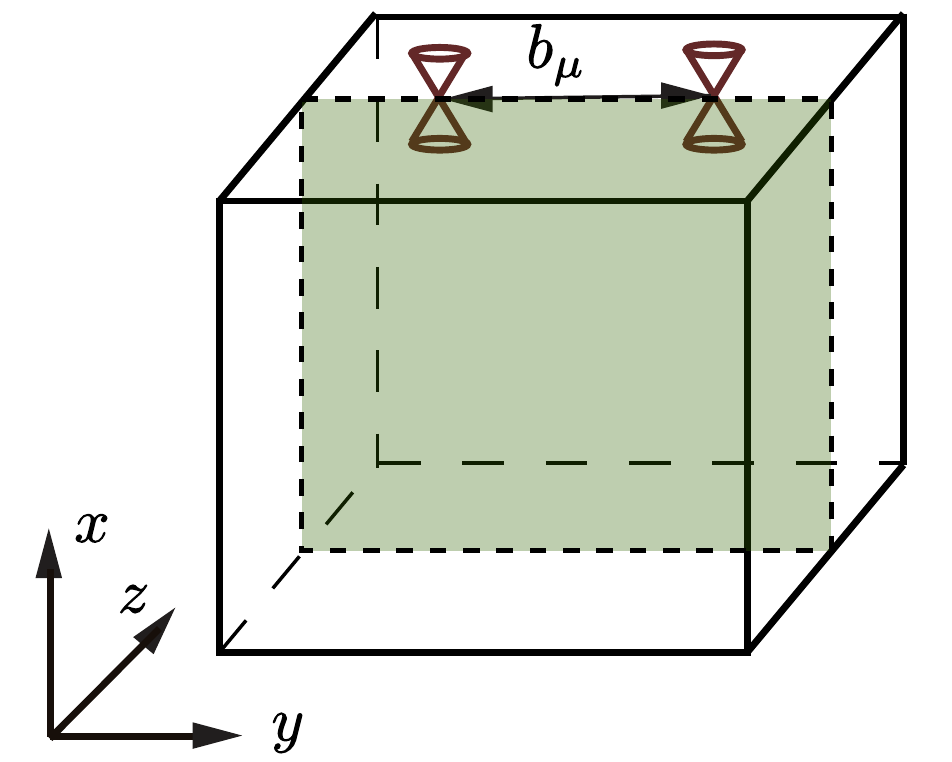}
   \caption{An illustration of a topological crystalline insulator in 3D with two surface Dirac cones localized on the surface perpendicular to the $x$-direction. $b_{\mu}=(b_{0},b_{y})$ is the energy-momentum separation of the Dirac nodes. For the model that we construct, the Dirac nodes are stabilized by $\mathcal{M}_{z}$ symmetry. We show the mirror plane in the bulk, and mirror lines on the surfaces by the green rectangle and its dashed boundary. }
\end{figure}

Three-dimensional mTCIs are characterized by  integer invariants: the mirror Chern numbers $C_{M}$\cite{teo2008}. 
To see the consequences of the $C_M$ let us consider a system with mirror symmetry $\mathcal{M}_{z}$ in the $z$-direction with $\mathcal{M}_{z}^{2}=-1.$ We can label eigenstates in the $k_{z}=0$ and $k_z=\pi$ planes of the Brillouin zone (BZ) with the eigenvalues $\pm i$ of $\mathcal{M}_{z},$ and this defines mirror Chern numbers $C_{M}(\Lambda)=\frac{C_{+i}(\Lambda)-C_{-i}(\Lambda)}{2},$ where $C_{\pm i}(\Lambda)$ is the usual Chern number of each mirror sector in the plane $\Lambda=0,\pi.$  When a $C_M$ is non-vanishing, then, on mirror-invariant surfaces, say one normal to $\hat{x},$ there will be Dirac cones  protected by the mirror symmetry. Furthermore, these cones lie in mirror invariant lines in the surface BZ projected from the corresponding $\Lambda$ planes. The number of \emph{stable} Dirac cones on each mirror line is given by $C_{M}(\Lambda)$\cite{teo2008}. If we allow for broken translation symmetry, then the total number of stable surface cones is $\mathcal{C}_{M}\equiv C_{M}(\Lambda_1)+C_{M}(\Lambda_2).$ We illustrate the case with $C_{M}(0)=2, C_{M}(\pi)=0$ in Fig.~\ref{fig:tciillustration} where we have two stable Dirac nodes on the surface perpendicular to $\hat{x}$ on the $k_{z}=0$ plane. 


In this article we will show that mTCIs have a robust electromagnetic (EM) response that is determined by both a topological property (the existence of stable surface states determined by $\mathcal{C}_{M}$), and a geometrical property (the momentum and energy locations of the surface nodes).  To show this, we first provide a lattice model for a mTCI  built from two copies of a 3D time-reversal invariant TI on a cubic lattice. By itself, this system has a trivial topological magnetoelectric effect, but when coupled to a field $b_\mu$ which preserves the mirror symmetry, yet splits the surface Dirac nodes in energy-momentum space, an additional EM response is generated. In this article we only consider systems which also retain $\mathcal{T}$ symmetry since the experimentally realized mTCIs have $\mathcal{T}$-symmetry, and it will simplify some discussions. 

For the simplest case with $\mathcal{C}_M=2,$ and with $\mathcal{T}$-symmetry, we can obtain a response theory of the mTCI via analogy with the 3D TI. In the continuum limit, the field $b_\mu$ in which we are interested couples to the theory precisely as a valley gauge field for the two species of surface Dirac cones/valleys.  {By performing a diagrammatic calculation in the continuum limit,} we find that the effective response action is given by:
\begin{align}
\mathcal {S}_{\rm TCI}[A,b]=\frac{e}{8\pi^{2}}\int \d^{4}x\,\epsilon^{\mu\nu\rho\sigma}\Theta f_{\mu\nu}F_{\rho\sigma},
\label{action}
\end{align} where $\Theta=\pi$ inside the bulk of the mTCI, and $f_{\mu\nu}=\partial_{\mu} b_{\nu}-\partial_{\nu}b_{\mu}$ is the field-strength of  $b_\mu$. The surface of the mTCI,  can be thought of as a domain wall of $\Theta$ from $\pi$ to $0$, and Eq. \eqref{action} implies a surface response $S_{2D}[A,b]=\frac{e}{4\pi}\int_{\rm surf}d^3x \epsilon^{\mu\nu\rho}b_{\mu}\partial_\nu A_\rho$  bound to the $\Theta$ domain wall. This surface response matches an EM response of a 2D Dirac semi-metal (DSM) with broken inversion symmetry if we identify $b_\mu$ with the energy/momentum separation of the Dirac node valleys\cite{TSMresponse,bernevigbook}. This is not so surprising, since the even number of  Dirac nodes on the surface of the mTCI is similar to the electronic structure of a 2D DSM.  However, we find precisely half the coefficient that would occur in a 2D DSM with mirror and $\mathcal{T}$ $(\mathcal{T}^2=-1)$ symmetries. 
Ultimately, the EM response of the mTCI implies localized charge and/or current density bound at defects in the $b_\mu$ field on the surface.
To verify the validity of the result Eq. \ref{action} obtained in the continuum limit, we explicitly calculate the microscopic origin of the response from a lattice model bound state calculation. 
Finally, we discuss experimental proposals and predictions.

\noindent\textit{Lattice Model for TCI.---}
Let us begin with the lattice Hamiltonian for a single copy of TI given by\cite{qhz2008}:
\begin{align}
H_{\rm TI}=&\sin k_{x}\Gamma^{x}+\sin k_{y}\Gamma^{y}+\sin k_{z}\Gamma^{z}\nonumber\\
&-(m+\cos k_{x}+\cos k_{y}+\cos k_{z})\Gamma^{0},
\label{ti}
\end{align}
where $m$ controls the bulk gap, and hence the topological phase. The matrices $\Gamma^\mu$ satisfy a Clifford algebra, and are given by $\Gamma^0=\tau^x s^0,~ \Gamma^x=\tau^y s^0,~\Gamma^y=\tau^zs^x,~\Gamma^z=\tau^zs^z,$ and $\Gamma^{5}=\tau^{z}s^{y}$, where the zeroth components $\tau^0$ and $s^0$ are identity matrices.
We can take $\tau$ to be an orbital degree of freedom and $s$ is spin; hence the time-reversal operator is $\mathcal{T}=is^{y}K$ where $K$ is complex conjugation.  Further, this model has mirror symmetries along the $i$-th directions with $\mathcal{M}_{i}=\Gamma^{i}\Gamma^{5}$ where $i=x,y,z$, and importantly $\mathcal{M}_{i}^2=-1.$ For example, we have $\mathcal{M}_{z} H_{\rm TI}(k_{x},k_{y},k_{z})\mathcal{M}_{z}^{-1}=H_{\rm TI}(k_{x},k_{y},-k_{z})$ 

To introduce a lattice model of the TCI, let us strictly enforce $\mathcal{M}_z,$ and add an additional flavor degree of freedom $\sigma^\mu$ to the TI model (\ref{ti}). We will  start with a block diagonal form,
\begin{align}
\label{eq:tciham}
H_{\rm TCI}^{(0)}=\sigma^0\otimes H_{\rm TI}.
\end{align} The topological phases and surface states of $H_{\rm TCI}^{(0)}$ are determined by $m$. Without loss generality, we consider a case where $-3<m<-1$; in this case there are two Dirac nodes (one for each copy) centered at the $\Gamma$-point on any surface (see SM). 
$\mathcal{T}$-symmetry enforces $C_{+i}(\Lambda)=-C_{-i}(\Lambda),$ and this model has $C_{M}(0)=2, C_{M}(\pi)=0,$ and $\mathcal{C}_{M}=2$. 

Various $\mathcal{M}_{z}$ preserving perturbations can be added to $H_{\rm TCI}^{(0)}$. 
Including some such perturbations we can write down a more generic lattice model for the TCI :
\begin{align}
H_{\rm TCI}&=\sin k_{x}\sigma^{0}\Gamma^{x}+(\sin k_{y}\sigma^{0}+b_y\sigma^y)\Gamma^{y}+\sin k_{z}\sigma^{0}\Gamma^{z}
\nonumber\\
&-(m+\cos k_{x}+\cos k_{y}+\cos k_{z})\sigma^{0}\Gamma^{0}+b_0\sigma^y,
\label{tci}
\end{align}
\noindent where tensor products are implicit and we will omit $\sigma^0$ from now on for compactness. One can verify that the Hamiltonian (\ref{tci}) is invariant under $\mathcal{M}_z$ and $\mathcal T$ (when $b_{0}=0$). One can also introduce $b_x$ and $b_z$ terms that couple to the Hamiltonian in a similar fashion to $b_y,$ and will fill out the entire $b_\mu=(b_0,b_x,b_y,b_z)$ field. Specific mirror symmetries will enforce some entries to be zero, for example $\mathcal{M}_{z}$ enforces $b_{z}=0$. 
We have left out a non-zero $b_{x}$, and some other possible $\mathcal{M}_z$ and $\mathcal{T}$-invariant terms since we will usually specialize to a particular surface ($\hat{n}=\hat{x}$) for convenience, and these additional terms, when small, will not impact our analysis.  We note that $b_{0}$ breaks time reversal, but not mirror,  and we include it in the Hamiltonian because it leads to an interesting EM response contribution.  As can be expected, we show in the SM that $b_0$ and $b_y$ move the Dirac nodes in the surface BZ to $(E,k_{y},k_{z})=(\pm b_{0},\pm b_{y},0),$ which is exactly what we need to generate a non-vanishing EM response.

\noindent\textit{Electromagnetic response.---}
First, we note that the topological magnetoelectric response of such a system, which is obtained by gapping the surface Dirac nodes with a $\mathcal{T}$-breaking mass,  is trivial since we have two copies of the usual TI. However, we now show that there is a response characteristic to to a mTCI, once a small $\mathcal{M}_{z}$ breaking mass term is added.

To this end, consider the TCI Hamiltonian \eqref{tci} in the continuum limit around the band inversion point $\Gamma$. At $b_0=b_y=0$, the continuum Hamiltonian has two identical copies, each given by
\begin{align}
H^{(a)}&=k_{x}\Gamma^{x}+k_{y}\Gamma^{y}+k_z\Gamma^{z}+m'\cos\theta^{(a)}\Gamma^{0}+m'\sin \theta^{(a)}\Gamma^{5}
\end{align}
where $m'>0$, $a=1,2$ corresponds to the two TI sectors with $\sigma^y=\pm1$, and we have included a generic $\theta^{(a)}$-angle for each block.
Under mirror symmetry $\mathcal{M}_{z}=\Gamma^{z}\Gamma^{5}$, the Hamiltonian satisfies:
\begin{align}
\mathcal{M}_{z}H^{(a)}(k_{x},k_{y},k_{z},\theta^{(a)})\mathcal{M}_{z}^{-1}=H^{(a)}(k_{x},k_{y},-k_{z},-\theta^{(a)}).
\end{align}
Thus, mirror symmetry enforces $\theta^{(a)}$ to take quantized values of $0$ or $\pi$. Indeed, our lattice model (\ref{tci}) maintains $\theta^{(a)}=\pi$ throughout the mTCI phase with $\mathcal{C}_M=2.$  

To calculate the continuum response we couple each of the continuum Hamiltonians with its own gauge field $A_{\mu}^{a}$ via $\mathbf{k}\rightarrow \mathbf{k}+\mathbf{A}^a$. A diagrammatic calculation similar to those in Refs.~\onlinecite{callanharvey,qhz2008} shows that
\begin{align}
S[A_{\mu}^{(a)}]=\frac{1}{32\pi^{2}}\int d^{4}x\, \theta^{(a)}(x)\epsilon^{\mu\nu\rho\sigma}F_{\mu\nu}^{(a)}F_{\rho\sigma}^{(a)}
\end{align} where $F_{\mu\nu}^{(a)}=\partial_{\mu}A_{\nu}^{(a)}-\partial_{\nu}A_{\mu}^{(a)}$ is the curvature associated with the gauge field $A_{\mu}^{(a)}$. The symmetric combination of the gauge fields $A_\mu^{(1)}$ and $A_\mu^{(2)}$ represents the usual EM field $A_{\mu},$ while the antisymmetric combination is a valley gauge field $b_{\mu}$, i.e., $eA_{\mu}=\tfrac{1}{2}(A_{\mu}^{(1)}+A_{\mu}^{(1)}), b_{\mu}=\tfrac{1}{2}(A_{\mu}^{(1)}-A_{\mu}^{(2)})$. Thus, the total effective action has a doubled/trivial topological magnetoelectric response, and a new mixed response given by
\begin{align}
\label{eq:tciresponse}
S[A,b]&=\frac{e}{8\pi^{2}}\int d^{4}x\,\Theta(x)\epsilon^{\mu\nu\rho\sigma}f_{\mu\nu}F_{\rho\sigma},
\end{align} 
where $\Theta=\theta^{(a)}=\theta^{(b)}=\pi$ and $f_{\mu\nu}=\partial_{\mu}b_{\nu}-\partial_{\nu}b_{\mu}$ is the curvature of the valley gauge field $b_{\mu}$. The same effective action can also be derived in a direct diagrammatic calculation by evaluating the diagram in  Fig.~2. Eq.\ \eqref{eq:tciresponse} is one of the main results of our paper. 

\begin{figure}[!h]
\label{fig:domainwall}
  \includegraphics[width=6cm]{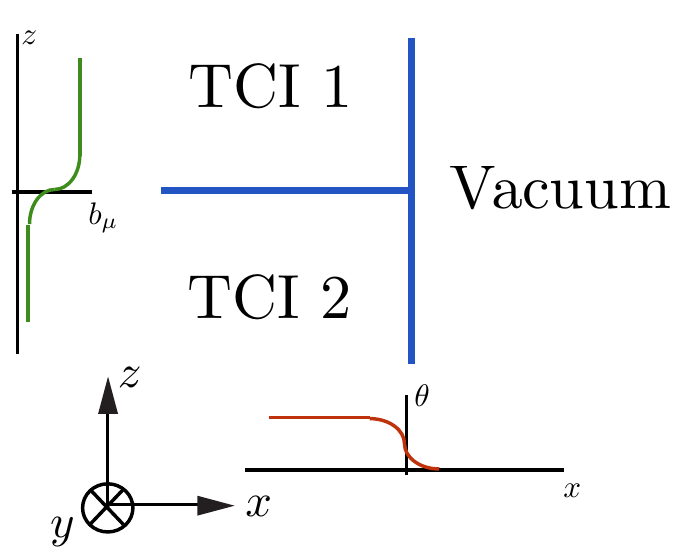}
   \caption{An illustration showing the kind of domain wall that can probe the response derived in Eq.~\ref{eq:tciresponse}. There is an interface between the TCI and the vacuum at $x=0$ and an interface between two TCIs with different $b_{\mu}$ at $z=0$. The quantities $b_{0},b_{y}$ naturally form a domain wall in the $x$ direction at $z=0$.}
\end{figure}

Let us illustrate the physical consequences of Eq. \ref{eq:tciresponse}. The surface of the mTCI can be thought of as a domain wall of $\Theta=\Theta(\vec x)$ where $\Theta$ changes from $\pi$ to 0 traversing from the mTCI to vacuum. Since $\Theta$ is only well-defined modulo $2\pi$, the sign of the response (\ref{eq:tciresponse}) is not fixed. To fix the sign
we need to  break \emph{mirror symmetry} at the surface which effectively chooses whether the domain wall steps up or down. This can be done by introducing a mass term $m_{A}\sigma^y\Gamma^5$, which breaks all $\mathcal{M}_i$, but preserves $\mathcal{T}.$ The effective action now reduces to a response localized at the $\Theta$ domain wall:  $S_{\rm TCI}=e\sgn(m_A)/(4\pi)\int_{\rm surf} d^3x \epsilon^{\mu\nu\rho}f_{\mu\nu}A_{\rho}$. Taking a derivative of the effective action with respect to $A_{\mu}$, we obtain 
the responses:
\begin{align}
j^0_{\rm surf}=-\frac{e\sgn{m_A}}{2\pi} \partial_z b_y,~~
j^y_{\rm surf}=\frac{e\sgn{m_A}}{2\pi} \partial_z b_0
\label{resp2}
\end{align}\noindent for static $b_y$ and $b_0.$
These two equations are analogs of the Streda formula and Ohm's law for a Hall current, where $\partial_z b_y$ and $\partial_z b_0$ are the $x$-component of the ``magnetic field" and the $z$-component of the ``electric field" of the 1-form $b_\mu$. 
The first equation of \eqref{resp2} indicates that  additional charge density is bound at a flux/vortex core of $\vec b$. For a domain wall $b_y=|b_y|\sgn z$ on the $yz$ surface of the TCI, which has a ``magnetic flux" of $\vec b$, Eq.\ \eqref{resp2} predicts that there exists a charge density of $j^0=e|b_y| \sgn(m_A)/\pi$ trapped at center of the domain wall. Macroscopically these responses arise from the half quantum Hall effect of each surface Dirac cone. They effectively see opposite electric and magnetic fields, but have opposite masses from $m_A.$ Hence, their responses add and do not cancel.
The defect structure that generates this response is illustrated for this case in Fig.\ \ref{fig:domainwall}. At the end we discuss the physical setup needed to experimentally probe this response. 

These results are based on the particular lattice model \eqref{tci}. However, they hold for any model with $\mathcal{C}_M=C_M(0)+C_M(\pi)=2$. We show in the SM that for a system with $\mathcal{M}_z$ and $\mathcal{T}$ (or even with weakly broken $\mathcal{T}$), $\mathcal{C}_M=2$ necessarily gives rise to two stable Dirac cones, and thus, upon the introduction of proper mirror-breaking mass terms, the response is described by Eq.\ \eqref{resp2}. 
For higher $\mathcal{C}_M=N$, there exist $N$ stable surface Dirac nodes, and in principle more complex TCI responses can be obtained (see Ref. \onlinecite{TSMresponse} for some related examples in 2D DSMs). 

For cases with $\mathcal{C}_M=0$, the surface Dirac cones can be gapped without breaking $\mathcal{M}_z$. However, if the bulk band crossing happens at different points in the bulk, then on certain surfaces the two Dirac cones can be located at different locations in the surface BZ. In this case gapping the Dirac cones, when mirror is preserved, requires breaking translational symmetry. By analogy with a weak TI\cite{FuKaneMele3D,MooreBalents07}, we dub the system with surface Dirac nodes protected by $\mathcal{M}_z$ and translational symmetry a weak mTCI. When translation symmetry is intact, a weak mTCI can have a response \eqref{resp2} on certain surfaces with mirror symmetry, but not necessarily all of them.

\noindent\textit{Microscopic origin of the response.---}
Eq.\ \eqref{resp2} is obtained from the continuum limit. However, from Eq.\ \eqref{tci} we see $b_y\sigma^y$ couples to the system like a gauge field {\it only} in the continuum limit, i.e., when $b_y$ is small. To obtain a complete picture, it is useful to verify the response from a microscopic calculation. 

From the TCI lattice model Eq.\ \eqref{tci}, we solve for the surface Dirac states on the boundary of a mTCI  ($x<0$) with the  vacuum ($x>0$). We then must solve for the bound states located at a domain wall of $b_y=|b_y|\sgn(z)$ on the surface $x=0$.  The surface Dirac cones are given by
\begin{align}
H_{2D}=&-(\sin k_y+b_y\sigma^y) s^x -\sin(k_z) s^z-m_A s^y\sigma^y,\nonumber\\
\label{2d}
\end{align}
where $m_A$ is the mirror symmetry breaking mass, and the momentum range over which $H_{2D}$ is valid is given by (see SM)
\begin{align}
|m-1+\cos k_y+\cos k_z|<1.
\label{range}
\end{align}\noindent In the continuum limit where $|b_y|$ is small, we can drop the sine in Eq.\ \eqref{2d} and neglect the upper cutoff for $k_{y,z}$. 
Next, for the non-uniform $b_y$ with a domain wall $b_y(z)=|b_y|\sgn z$,  the solution of Eq.\ \eqref{2d} is given by\begin{align}
\Psi(x=0,k_y,z)=&\exp\left\{-\int_0^z [k_y\sigma^y+b_y(z')]dz'\right\}\Psi_0(k_y),
\label{1d}
\end{align}
where $\Psi_0(k_y)$ satisfies $s^y\sigma^y\Psi_{0}(k_{y})=-1$.
Eq.\ \eqref{1d} describes {\it two} bound states at each $k_y$,  corresponding to the two eigenvalues of $\sigma^y$, both localized in $z$ direction at the zero of the integrand of the exponent. Since $b_y$ ranges from $-|b_y|$ to $|b_y|$, only  states for which $|k_y|<|b_y|$ has bound state solutions. Therefore, the total number of bound states is $2\times 2|b_y|/(2\pi/L_y)$, where the first prefactor of 2 corresponds to the two eigenstates of $\sigma^y$ that form a Kramers' pair. For a finite system, on the same surface $x=0$ there exist an opposite domain wall with $b_y(z)=-|b_y|\sgn z$. There exist same number of bound states at the opposite domain wall domain wall, only with $s^y\sigma^y=-1$.
 When $m_A\to 0$, these states all have zero energy, but  when a small but finite $m_A$ is introduced, these states localized at opposite domain walls are lifted from zero energy and can be unambiguously filled. Due to the usual arguments~\cite{jackiw1976s}, each state generates a localized charge $-\frac {e} 2\sgn m_A$. Therefore we find that the bound state charge density is $j^0=-e|b_y| \sgn(m_A)/\pi$, and is in agreement with the result from previous Eq. \eqref{resp2}. We note that to see this response, we only need to break mirror symmetry with an infinitesimal mass term, while time reversal symmetry is intact.

For a larger magnitude of the domain wall $\lvert b_y\rvert$, the charge response can  deviate from the prediction from Eq.\ \eqref{resp2} and become non-universal, but this happens simultaneously with a gap closing transition in the bulk. Whether the charge response is universal is tied to the fate of the surface Dirac nodes in Eq.\ \eqref{2d}. For a sufficiently small $|b_y|$, the Dirac nodes simply get shifted. However,  depending on the momentum range given by \eqref{range}, a larger value of $|b_y|$ can either eliminate the Dirac nodes or introduce additional Dirac nodes that can gap out each original one. In both cases, there is a gap closing in the bulk which indicates a transition from a mTCI to a trivial insulator. We show in the SM that, as long as the bulk gap does not close, the response \eqref{resp2} from the continuum model remains valid, {\it even if\,}  $b_y$  does not precisely couple like a valley gauge field in the lattice model.
However for the cases when a large $b_y$ eliminates or cancels the original Dirac nodes on the two sides of the domain wall, the charge density bound at the domain wall becomes non-universal and only depends on the upper cutoff in $k_y$ given by Eq.\ \eqref{range}. 



\noindent\textit{Implication for experiments.---}
Generically, as shown in Ref.~\onlinecite{tang2014}, the Dirac nodes on the surface of a TCI can be moved around by symmetry preserving compression/dilation or uniaxial stretching. In the specific case of $\textrm{SnTe}$, the surface Dirac nodes perpendicular to the (001) direction arise at $\pm \mathbf{k}_{1},\pm \mathbf{k}_{2}$ and are protected by mirror symmetry along $(110)$ and $(1\overline{1}0)$ axes respectively. Compression or dilation being isotropic moves all the surface Dirac nodes in or out meaning $b_{\mu}$ for both pairs of Dirac nodes increases or decreases. If the system has $C_{4}$ symmetry additionally, as in \textrm{SnTe}, then the uniaxial stretching breaks $C_{4}$ symmetry and the Dirac nodes at $\pm \mathbf{k}_{1}$ move out ($b_{\mu,1}$ increases) while the nodes at $\pm \mathbf{k}_{2}$ move in ($b_{\mu,2}$ decreases) or vice versa. The way to generate non-zero masses for these surface Dirac fermions as discussed in Ref.~\onlinecite{hsiehtci} is through structural distortions where the atoms in the underlying lattice are displaced by $\mathbf{u}$. The mass term induced by such a structural distortion was shown to be $m_{j}\propto (\mathbf{u}\times \mathbf{K}_{j})\cdot\hat{n}$ where $\mathbf{K}_{j}$ is the momentum location of the Dirac node on the surface perpendicular to $\hat{n}$. We expect that the structural distortion leading to a ferroelectric phase as described in Ref.~\onlinecite{hsiehtci} would lead to domain wall charge/currents described in this paper.


\acknowledgements{
We acknowledge useful discussions with Onkar Parrikar, Vatsal Dwivedi, Awadhesh Narayan and Victor Chua.  STR and TLH are supported by the ONR YIP Award N00014-15-1-2383. YW is supported by the Gordon and Betty Moore Foundation's EPiQS Initiative through Grant No. GBMF4305 at the University of Illinois.}

\onecolumngrid

\newpage

\renewcommand{\theequation}{S\arabic{equation}}
\renewcommand{\thefigure}{S\arabic{figure}}
\renewcommand{\bibnumfmt}[1]{[S#1]}
\renewcommand{\citenumfont}[1]{S#1}

\centerline{\large \textbf{Supplemental Material}}

\section{I.~~~Diagrammatic calculation of response}
\begin{figure}[h]
\includegraphics[width=0.4\columnwidth]{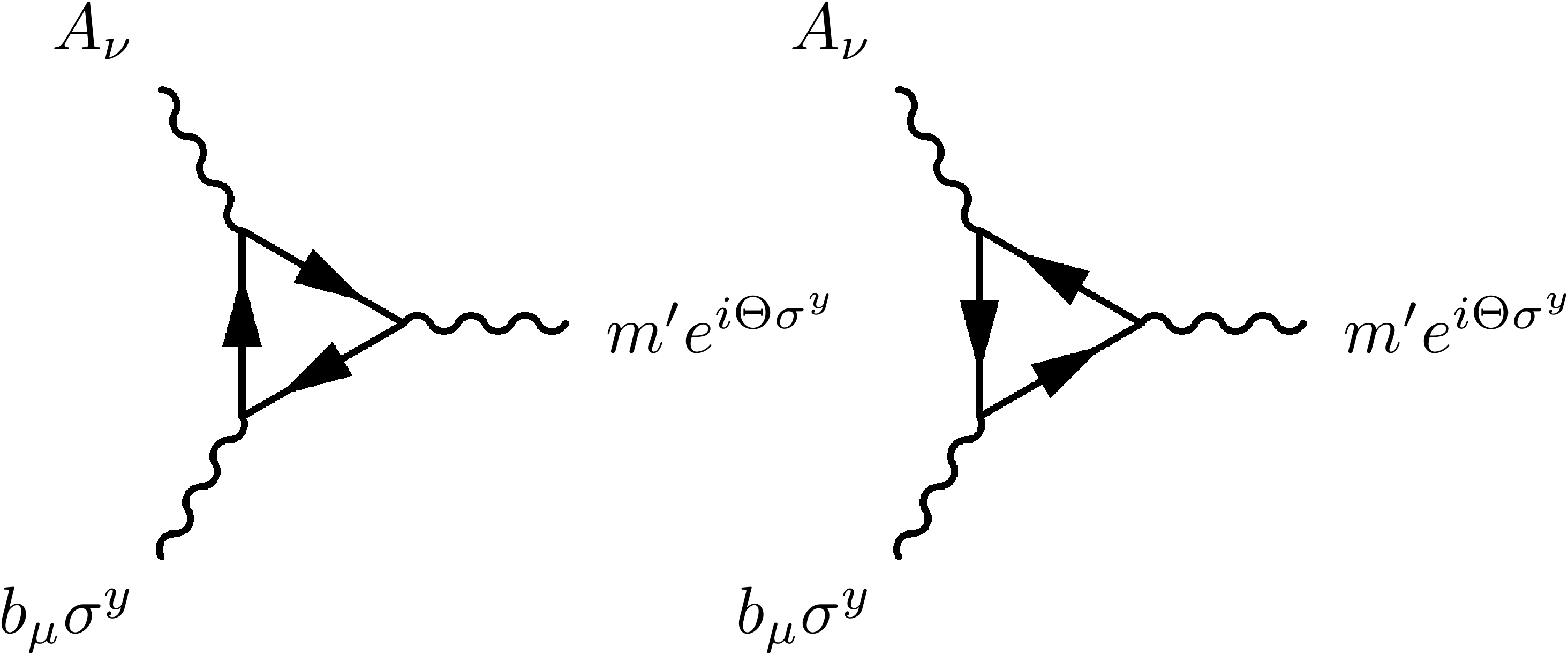}
\caption{Relevant Feynman diagrams for the effective action indicated in Eq.~\eqref{sm_eq:sm1}.} 
\label{sm_fig:feyndiag}
\end{figure}

In this section we provide a direct diagrammatic calculation for the response Eq.\ (8) in the main text.  We begin with the continuum version of the Hamiltonian (4) in the main text,
\be
H_{\rm TCI}=k_x \Gamma^x + (k_y+b_y\sigma^y)\Gamma^y+k_z\Gamma^z+m'\cos\Theta\Gamma^0+m'\sin\Theta\sigma^y\Gamma^5,
\label{sm_sm:1}
\ee
where tensor products are implicit and $\sigma^0$'s have been omitted.
The relevant diagrams that need to be calculated are shown in Fig.~\ref{sm_fig:feyndiag}. 

The two diagrams are equal due to symmetry and contribute an extra factor of 2. Evaluating the diagrams as in Ref.~\onlinecite{callanharvey_s}, we find:
\begin{align}
S[A,b]=&-2\times\int d^4x\frac{e^2}{16\pi^2}\epsilon^{\mu\nu\rho\sigma}\Tr_\sigma[\partial_\mu(\Theta\sigma^y)(b_\nu\sigma^y)\partial_{\rho}A_{\sigma}]
\end{align}
After tracing over $\sigma$ and integrating by parts, we obtain
\be
\label{sm_eq:sm1}
S[A,b]=\frac{e}{8\pi^2}\int d^{4}x\,\Theta(x)\,\epsilon^{\mu\nu\rho\sigma}f_{\mu\nu}F_{\rho\sigma},
\ee
which is Eq.\ (8) in the main text.

\section{II.~~~Stability of the surface Dirac nodes for $\mathcal{C}_{M}=2$}

In this section, we show that for a mTCI with mirror symmetry (say along z direction, $\mathcal{M}_z$) and time-reversal symmetry ($\mathcal{T}$), the stability of the surface Dirac nodes is related to ${\mathcal C}_M$. Particularly, we prove that for the case of $\mathcal{C}_M=2$, there are two stable Dirac nodes on {\it all} mirror symmetric surfaces. Upon including proper mirror-breaking mass terms and a defect in the $b$ field, the TCI response we postulate can be obtained.

Our discussion begins with a heuristic analysis of a TCI that consists of two blocks of a strong TI that has inversion symmetry $\mathcal{I}$ in addition to time-reversal and mirror symmetry. Each copy contributes $\mathcal{C}_{M}=1$ to the mirror Chern number. In the presence of inversion symmetry, the mirror Chern number for each of the TI copies $\mathcal{C}_M$ can be directly related to the mirror chirality $\eta$ defined\cite{teo2008_s}  at TR invariant points on the mirror planes. Using the mirror chirality, we show that for all cases where $\mathcal{C}_M=2$, the TCI supports two stable surface Dirac nodes.

In the second subsection, we provide a more general proof that for {\it all} cases with $\mathcal{M}_z$ and $\mathcal{T}$, $\mathcal{C}_M=2$ necessarily indicates two stable Dirac cones on the surfaces. All of these arguments are a slightly expanded version of what is present in Ref.~\onlinecite{teo2008_s}.

\subsection{A.~~~TCI from two strong TI's}
Let us consider our 3D lattice model for a TCI which consists of two blocks of the usual 3D strong TI. The model for the 3D TI is
\begin{align}
H_{TI}=v_x\sin k_x \Gamma^x+v_y\sin k_y \Gamma^y +v_z\sin k_z \Gamma^z -(M+\cos k_x+\cos k_y+\cos k_z)\Gamma^0
\end{align}\noindent where $\Gamma^x=\tau^y\otimes\mathbb{I}, \Gamma^y=\tau^z\otimes s^x, \Gamma^z=\tau^z\otimes s^z, \Gamma^0=\tau^x\otimes\mathbb{I}$ and $\Gamma^5=\tau^z\otimes s^y.$ The mirror operators according to our paper are $\mathcal{M}_i=\Gamma^5\Gamma^i$ which satisfy $\mathcal{M}_{i}^2=-1.$ In particular $\mathcal{M}_z=i\mathbb{I}\otimes s^x.$

Given this model we can construct a TCI Hamiltonian by taking two blocks of $H_{TI}$, although we will see that it is important to consider generic velocities in each block. To simplify the analysis, we will choose each velocity $v_i$ for both blocks to be proportional to a positive constant $v$ which we will set to $1$ for convenience. Then each velocity parameter represents just the sign of the velocity, i.e., $v_i=\pm 1.$ Given a choice of mirror operator we can define the mirror chirality of the $k\cdot P$ Hamiltonians around each time-reversal invariant momentum. For our simple model, if we choose mirror operator $\mathcal{M}_{i}$ then the mirror chirality for each $4\times 4$ Dirac block is 
\be
\chi_M={\rm{sgn}}({v_j v_k}),
\ee
 where $i\neq j \neq k$ as derived in Ref.~\onlinecite{teo2008_s}.

At this point, we concern ourselves with the mirror symmetry $\mathcal{M}_z$ and consider surface Hamiltonians perpendicular to $\hat{x},\hat{y}$. For $H_{TI}$ if we choose $M=0$ the bands at $\Gamma$ are inverted and will lead to surface states. For a surface with normal vector $\hat{x}$ we find the surface Hamiltonian
\begin{equation}
H_{{\rm surf},x}=({\rm{sgn}}v_x)(v_y\sin k_y s^x+v_z\sin k_z s^z)
\end{equation}\noindent and the mirror operator projects to $is^x$ on the surface. Note that the dependence on the sign of $v_x$ 
arises from the bound state condition where that sign chooses which eigenvalue of $\Gamma^x\Gamma^0$ corresponds to which surface. If we switch the sign then the bound state on the $+\hat{x}$ surface corresponds to the opposite eigenvalue and the projected surface matrices each pick up a negative sign. 

For the $+\hat{y}$ direction we have
\begin{equation}
H_{{\rm surf},y}=({\rm{sgn}} v_y)(v_x\sin k_x s^z+v_z\sin k_z s^x)
\end{equation} and, importantly,  the mirror operator projects to $is^z$ on this surface. Both of these surface Hamiltonians are protected as long as mirror symmetry is preserved.

Now if we add a second block to represent a TCI we have some choices of velocity signs. In fact, we have a sign choice for each velocity. There are $2^3=8$ choices, but, they only give rise to four distinct Hamiltonians as far as the stability analysis is concerned. These are:
\begin{eqnarray}
H_{{\rm surf},x,1}&=&\sin k_y \mathbb{I}\otimes s^x+\sin k_z \mathbb{I}\otimes s^z\\
H_{{\rm surf},x,2}&=&\sin k_y \mathbb{I}\otimes s^x+\sin k_z \mu^z\otimes s^z\\
H_{{\rm surf},x,3}&=&\sin k_y \mu^z\otimes s^x+\sin k_z \mathbb{I}\otimes s^z\\
H_{{\rm surf},x,4}&=&\sin k_y \mu^z\otimes s^x+\sin k_z \mu^z\otimes s^z
\end{eqnarray}\noindent and the other four differ from these by a global sign multiplying the full Hamiltonian. In cases $H_{{\rm surf},x,3}$ and $H_{{\rm surf},x,4}$ we can find mass terms which preserve $\mathcal{M}_z$, e.g. $\mu^x\otimes s^x$ and $\mu^x\otimes \mathbb{I}$ respectively. 

We can do something similar for the other surface type
\begin{eqnarray}
H_{{\rm surf},y,1}&=&\sin k_x \mathbb{I}\otimes s^z+\sin k_z \mathbb{I}\otimes s^x\\
H_{{\rm surf},y,2}&=&\sin k_x \mathbb{I}\otimes s^z+\sin k_z \mu^z\otimes s^x\\
H_{{\rm surf},y,3}&=&\sin k_x \mu^z\otimes s^z+\sin k_z \mathbb{I}\otimes s^x\\
H_{{\rm surf},y,4}&=&\sin k_x \mu^z\otimes s^z+\sin k_z \mu^z\otimes s^x.
\end{eqnarray}\noindent In this case as well, the last two Hamiltonians are unstable even when mirror is preserved. 

Interestingly, we see that the
two Dirac nodes
 in  $H_{{\rm surf},x/y,3}$ have the opposite helicity, but those in $H_{{\rm surf},x/y,4}$ have the same helicity, and yet they can be gapped in either case. Hence, the helicity is not what we should be using to characterize the surface states. However, one can easily check that if the \emph{mirror chirality} $\chi$ between the blocks are the same (different), then the resulting surface states are stable (unstable). As shown in Teo, Fu, and Kane~\cite{teo2008_s}, the mirror chirality determines the sign of the \emph{change} in mirror Chern number when there is a band inversion. Thus, we can correlate the cases with stable surface states as having $\mathcal{C}_M=\pm 2$ and in the unstable cases $\mathcal{C}_M=0.$

Let us look at another example. Consider $H_{{\rm surf},x,1}$ and $H_{{\rm surf},x,3}$ and turn on the $b_y$ shift generated by $b_y \mu^y\otimes \Gamma^y$ in the bulk Hamiltonian. We immediately find that this perturbation shifts the Dirac nodes in $H_{{\rm surf},x,1}$  but gaps the Dirac nodes in $H_{{\rm surf},x,3}.$ Hence, if we have a domain wall in $b_y$, this term will gap out the Dirac nodes in $H_{{\rm surf},x,3}$ instead of shifting them. We can see that it is crucial to have the $\mu^z$ term in the surface Hamiltonian which will anti-commute with the $\mu^y$ term in the shift. We could have also chosen to shift with $\mu^x$ but the same result applies. 
\subsection{B.~~~General proof}
Let us consider a system with mirror symmetry along $z$ direction ($\mathcal{M}_z$) and time reversal symmetry ($\mathcal{T}$), characterized by a mirror Chern number
\be\label{sm_sm:2}
\mathcal{C}_M\equiv C_M(k_z=0)+C_M(k_z=\pi)=2,
\ee
where $k_z=0$ and $k_z=\pi$ are two invariant momenta under mirror symmetry. We claim that there necessarily exist two Dirac cones on a $x$-surface that are protected by $\mathcal{M}_z$. 

Before we start, we note that due to $\mathcal{M}_z$ symmetry, Dirac cones located at $k_z\neq 0,\pi$ always appear in pairs with opposite $k_z$ values. These Dirac cones can generically gap out each other when translational symmetry along $z$ is broken, i.e., at the domain wall of $m_A(z)$ we consider in the main text. Therefore, such Dirac cones do not have nontrivial contribution to the topological response we consider, and we will focus only at   $k_z=0,\pi$ planes.

At $k_z=0,\pi$, $\mathcal{M}_z$ is a good quantum number, and we can divide this 2D system into two subsystem with $\mathcal{M}_z=\pm i$. From the definition $\mathcal{C}_M=(\mathcal{C}_i-\mathcal{C}_{-i})/2$ and by time-reversal symmetry, we have $\mathcal{C}_i=2$ and $\mathcal{C}_{-i}=-2$.

The fact that $\mathcal{C}_i=2$ indicates two chiral modes in the $+i$ sector at the edge of the $xy$ plane with at $k_z=0$ and/or $k_z=\pi$. Hence at an $x$-surface, to linear order,
\be
H_{i, k_z=0,\pi}=v_y k_y \mu^0,
\ee
where $\mu^0$ is a 2-by-2 matrix corresponding to the two edge modes. In principle there is nothing enforcing the $v_y$'s to be the same for the two chiral modes, but they should have the same sign. Generally, it can be shown that a different magnitude of $|v_y|$'s would lead to no change to the final conclusion. 

On the other hand, $C_{-i}=-2$ also indicates two edge modes in the $-i$ sector that are related to $+i$ sector by time-reversal symmetry, hence $H_{-i, k_z=0,\pi}=-v_y k_y \mu^0$. Combining $\pm i$ sectors, 
\be
H_{k_z=0,\pi}=v_y k_y s^x\otimes\mu^0,
\ee
where we define $s^x=\pm 1$ for $\mathcal{M}_z=\pm i$. Thus, in the subspace of the four surface bands, the form of the mirror operators is given by $\mathcal{M}_z=is^x\otimes\mu^0$.

We can now introduce the $k_z$ dependence back. Other than a constant term, mirror symmetry $\mathcal{M}_z$ enforces that the only allowed terms are of the form $\sim v_z k_z s^{y,z} \otimes \mu^{0,1,2,3}$.
 In general the 2D Hamiltonian at the $x$-surface is
\be
H_{yz}=v_y k_y s^x\otimes\mu^0+v_z k_z \Pi_z,
\ee
where $\{{\mathcal{M}_z},\Pi_z\}=0$ which arises from the fact that $\mathcal{M}_{z}H_{yz}(k_{y},k_{z})\mathcal{M}_{z}^{-1}=H(k_{y},-k_{z})$ due to mirror symmetry. 
This Hamiltonian corresponds to two Dirac cones.\footnote{Due to the anti-commutation with $\mathcal{M}_z$, the four eigenvalues of $\Pi_z$ are necessarily of the form $(+a,-a,+b,-b)$, where the two states with $\Pi_z\ket{\alpha}=a\ket{\alpha}$ and  $\Pi_z\ket{-\alpha}=-a\ket{-\alpha}$ are related by $\ket{-\alpha}=-i\mathcal{M}_z\ket{\alpha}$. In each block, say the one with eigenvalues $\pm a$, $\Pi_z$ projects to $a \sigma^z$ and $s^x\otimes \mu^0\equiv -i\mathcal{M}_z$ projects to $\sigma^x$. The reduced Hamiltonian in that block is $H=v_y k_y \sigma^x + a v_zk_z \sigma^z$, hence the Dirac cone.}
 {\it Any} gapping term necessarily involves $s^{y,z}$, which is forbidden by $\mathcal{M}_z$.

  \section{III.~~~Microscopic origin of the response}
\subsection{A.~~~Surface states of a TCI}\label{sm_sec:domwall}
We derive the surface Dirac states from the explicit tight binding model for the TCI. We start with the lattice model used in the main text with $b_y$ turned on:
\begin{align}
\label{sm_eq:tti}
H_{TCI}\lbrack\mathbf{k},\theta\rbrack&=\sin k_{x}\Gamma^{x}+(\sin k_{y}+b_y\sigma^y)\Gamma^{y}+\sin k_{z}\Gamma^{z}
-(m+c\cos k_{x}+c\cos k_{y}+c\cos k_{z}
+c\cos\theta)\Gamma^{0}+\sin \theta\,\Gamma^{5}
\end{align}
where $c$ is a constant that we set to $1$ and $\theta=\pi$ is the adiabatic parameter field. Let us remind ourselves of the Dirac $\Gamma$ matrix basis that we choose:
\begin{align}
\Gamma^0=\tau^x s^0,~ \Gamma^x=\tau^y s^0,~\Gamma^y=\tau^zs^x,~\Gamma^z=\tau^zs^z,~\Gamma^5=\tau^z s^y.
\end{align}

We solve for the states bound at the two $yz$ surfaces perpendicular to the $x$ direction.
Let us first consider the TCI Hamiltonian \eqref{sm_eq:tti} without the terms involving $\Gamma^{y},\Gamma^{z}$, and solve for the {\it zero energy} eigenstates. Since the $x$-direction is no longer periodic in this case, we explicitly return to real space in $x$ direction. The reduced Hamiltonian is
\begin{align}
H_1=\sum_{x=1}^{N-1} \left\lbrack\frac{i}{2}\Gamma^{x}(c^{\dagger}_{x+1}c_{x}-c_{x}^\dagger c_{x+1})+\frac{1}{2}\Gamma^{0}(c^{\dagger}_{x+1}c_{x}+c_{x}^\dagger c_{x+1})\right\rbrack - \sum_{x=1}^N (m-1+\cos k_{y}+\cos k_{z})\Gamma^{0}c^\dagger_x c_x,
\label{sm_eqh1}
\end{align}
where we have suppressed the $k_y$ and $k_z$ indices in $c_x$ operators.
We use the ansatz 
\be
|\Phi\rangle=\sum_{x'=1}^N \lambda^{x'} c^\dagger_{x'}|0\rangle
\ee
for the wave function. Substituting this ansatz for the edge state, we have:
\begin{align}
\Gamma^0H|\Phi\rangle=&\left[\frac{i}{2}\Gamma^0\Gamma^x\left(\frac{1}{\lambda}-\lambda \right)+\frac{1}{2}
\left(\frac{1}{\lambda}+\lambda \right)-(m-1+\cos k_y+\cos k_z) \right]|\Phi\rangle\nonumber\\
&-\frac{i}{2}\Gamma^0\Gamma^x(c_1^\dagger-\lambda^{N+1}c_N^\dagger)|0\rangle -\frac{1}{2}(c_1^\dagger+\lambda^{N+1}c_N^\dagger)|0\rangle =0.
\label{sm_17}
\end{align}
We can take $i\Gamma^0\Gamma^x=-\tau^z=\pm 1$, and then the first line reduces to an algebraic equation. However, we need to be careful about the two end terms at $x=1$ and $x=N$ in the second line. For the state with $i\Gamma^0\Gamma^x=-1$, i.e. $\tau^z=1$, $c_1^\dagger$ cancels out, and $c_N^\dagger$ term can only be neglected if $|\lambda|<1$. For the same reason if $i\Gamma^0\Gamma^x=1$, i.e. $\tau^z=-1$, we need to impose $|\lambda|>1$. Therefore $\tau^z=1$ corresponds to the ``left" $yz$ surface of the TCI
 and $\tau^z=-1$ corresponds to the ``right" $yz$ TCI surface. Solving for the first line of Eq.\ \eqref{sm_17} with $\tau^z=-i\Gamma^0\Gamma^x=\pm1$ we find $\lambda=(m-1+\cos k_y+\cos k_z)^{\pm1}$. It is easy to see that, for both surface states corresponding to $\tau^z=\pm 1$, it is required that 
 \begin{align}
 |m-1+\cos k_y+\cos k_z|<1.
 \label{sm_range}
 \end{align}

We can now put back the $\Gamma^y=\tau^z s^x$ and $\Gamma^z=\tau^z s^z$ terms, namely,
\begin{align}
H_2=(\sin k_{y}+b_y\sigma^y)\Gamma^{y}+\sin k_{z}\Gamma^{z}
\label{sm_eqh2}
\end{align}
 into the TCI Hamiltonian. Since $\tau^z$ commutes with both Gamma matrices, we can simply substitute it with the corresponding eigenvalue. Focusing on the right surface where $\tau^z=-1$, we obtain the surface dispersion 
\be
H_{2D}(k_y,k_z)=-(\sin k_y+b_y\sigma^y) s^x- \sin(k_z) s^z,
\label{sm_h2d}
\ee\noindent
with $k_{y,z}$ satisfying $|m-1+\cos k_y+\cos k_z|<1$. Eqs.\ \eqref{sm_range} and \eqref{sm_h2d} are Eqs.\ (11) and (10) in the main text. 

For $-2<m<0$, the range \eqref{sm_range} for $k_y$ and $k_z$ are centered around the $\Gamma$ point. Particularly for $k_z=0$ the range for $k_y$ is given by
\begin{align}
-\Lambda_m<k_y<\Lambda_m,~~\Lambda_m=\arccos(-1-m).
\label{sm_range55}
\end{align}
For simplicity, throughout the rest of the Supplemental Material we focus on this case. However, we have verified that our final conclusion will hold for other values of $m$. For $2<m<4$ the range for $k_y$ and $k_z$ is centered around $(\pi,\pi)$, and for $0<m<2$ the range for $k_y$ and $k_z$ is centered around $(\pi,0)$ and $(0,\pi)$; our analysis around $(0,0)$ point can be easily carried over to these cases.

In general the 2D Hamiltonian in \eqref{sm_h2d} describes the surface Dirac nodes of the TCI. For a small $b_y$, the Dirac nodes are simply shifted from $\Gamma$ point. However, a larger $b_y$ can either shift the original Dirac nodes outside the validity range, or introduce additional Dirac nodes with opposite helicities compared to original ones and then annihilate them. In both cases the transition happens when a Dirac node is at the upper limit of the range \eqref{sm_range55}, i.e.  $k_y=\pm \Lambda_m$. This corresponds to 
\begin{align}
|b_y^{cr}|=\sin(\Lambda_m),
\label{sm_cr}
\end{align}
and for $-2<m<0$ we have $|b_y^{cr}|=\sqrt{-m(m+2)}$.

 \subsection{B.~~~Domain wall states on the surface of TCI}\label{sm_sec:surfdomwall}
In this section we analyze the surface states of the TCI bound at domain walls of $b_y$ in the $z$-direction.
The domain walls we consider are mirror symmetric, given by 
\begin{align}
b_y(z)=\begin{cases}|b_y^0|,&~~z> z_0,\\
-|b_y^0|,&~~|z|<z_0\\
|b_y^0|,&~~z<- z_0.\end{cases}
\end{align}
Since the two domain walls are spatially separated, it suffices to consider each of them separately. After a simple coordinate shift, we focus on a domain wall where $b_y(z)=|b_y^0|$ as $z\to\infty$ and $b_y(z)=-|b_y^0|$ as $z\to-\infty$.
To do this the most intuitive way would be start from the surface Dirac Hamiltonian \eqref{sm_h2d}, and replace $b_y$ with $b_y(z)$ and $k_z$ with $-i\partial_z$. However, with this method the role of the momentum range of the surface states \eqref{sm_range} would become unclear. In this section we instead directly solve for the domain wall bound state from the bulk Hamiltonian. The derivation follows similar procedures as in the previous section.

Similar to the previous Section, we split the TCI Hamiltonian into two parts, $H_{TCI}=H_1+H_2$, where $H_1$ and $H_2$ are defined similarly as in the previous section, only with the $z$-direction expressed in real space. Explicitly, we have:
\begin{align}
H_1=\sum_{x,k_y,z} &\left\lbrack\frac{i}{2}\Gamma^{x}(c^{\dagger}_{x+1,k_y,z}c_{x,k_y,z}-c_{x,k_y,z}^\dagger c_{x+1,k_y,z})+\frac{1}{2}\Gamma^{0}(c^{\dagger}_{x+1,k_y,z}c_{x,k_y,z}+c_{x,k_y,z}^\dagger c_{x+1,k_y,z}) \right.\nonumber\\
&\left.-(m-1+\cos k_{y})\Gamma^{0}c^\dagger_{x,k_y,z} c_{x,k_y,z}-\frac{1}{2}\Gamma^0(c^{\dagger}_{x,k_y,z+1}c_{x,k_y,z}+c_{x,k_y,z}^\dagger c_{x,k_y,z+1}),\right\rbrack \nonumber\\
H_2=\sum_{x,k_y,z} &\left \lbrack\sin (k_{y}+b_y(z)\sigma^y)\Gamma^{y}c^\dagger_{x,k_y,z} c_{x,k_y,z}+\frac{i}{2}\Gamma^{z}(c^{\dagger}_{x,k_y,z+1}c_{x,k_y,z}-c_{x,k_y,z}^\dagger c_{x,k_y,z+1})+m_A\sigma^y\Gamma^{5}c^\dagger_{x,k_y,z} c_{x,k_y,z}\right],
\end{align}
where we have added a small mirror symmetry breaking mass $m_A$ to $H_2$.  We use the ansatz for the domain wall state given by
\be
|\Phi(k_y)\rangle=\sum_{x,k_y,z=1}^N \lambda(z)^x \exp^{-\sum_{z'=1}^z f(z')} c^\dagger_{x,k_y,k_y,z}|0\rangle,
\label{sm_asz}
\ee
where $f(z)$ is a function that we will relate to $b_y(z)$ later. We emphasize that, this ansatz describes a bound state {\it only} if $\lambda(z)\neq 1$, {\it and} the  function $f(z)$ goes through zero with a {\it positive} slope.

We use the ansatz to solve for a zero energy state of $H_{1}$. Similar to the first line of Eq.\ (\ref{sm_17}), we have for the zero energy solution
\begin{align}
&\frac{i}{2}\Gamma^0\Gamma^x\left[\frac{1}{{\lambda (z)}}-{\lambda (z)}\right]+\frac{1}{2}
\left[\frac{1}{{\lambda (z)}}+{\lambda (z)}\right]-\left[m-1+\cos k_y+\frac{\exp [f(z+1)]+\exp [-f(z-1)]}2\right]\nonumber\\
\approx&\frac{i}{2}\Gamma^0\Gamma^x\left[\frac{1}{{\lambda (z)}}-{\lambda (z)}\right]+\frac{1}{2}
\left[\frac{1}{{\lambda (z)}}+{\lambda (z)}\right]-\left[m-1+\cos k_y+\cosh f(z)\right]=0.
\label{sm_20}
\end{align}
where in the second step we have assumed that $f(z)$ is a slow varying function.
Just like the case of the surface state, for this equation to make sense the condition is that for $i\Gamma^0\Gamma^x=-\tau^z=\pm 1$, $|\lambda(z)| \gtrless 1$. Substituting this into the solution of Eq.\ (\ref{sm_20}),
\begin{align}|m-1+\cos k_y+\cosh f(z)|<1.
\label{sm_range2}
\end{align}

We still need to determine the form of $f(z)$. This can be done by solving for eigenstates of $H_2$, which involves $\Gamma^y$ and $\Gamma^z$.  We have for $\tau^z=-1$
\be
\label{sm_eqh2'}
H_2=-[\sin k_y+b_y(z)\sigma^y] s^x- i\sinh f(z) s^z - m_As^y\sigma^y.
\ee\noindent
This Hamiltonian is solved by 
\be
\label{sm_fz}
f(z)=-\sinh^{-1}\{[\sin k_{y}\sigma^y+b_y(z)]s^y\sigma^y\},\,\,\sigma^{y}s^{y}=-1,
\ee
where $\sinh^{-1}$ is meant to act separately on each eigenstate of $\sigma^y$ and $s^y$.
In order for this to be a bound state, the slope of $f(z)$ has to be positive, which enforces $s^y\sigma^y=-1$. Therefore, the energy of this state is $E=m_A$. The function $f(z)$ goes through zero where $\sin k_y \sigma^y+b_y(z)=0$, and therefore there are two bound states localized at $z=\pm z_0$, where {$\sin k_y=b_y(z_0)$ and
the $\pm$ corresponds to the two eigenvalues of $\sigma^y$. The doubling of the bound states corresponds to the fact that our particular TCI model are composed of two copies of the TI. 

 Eq.\ \eqref{sm_range2} becomes
\begin{align}
\label{sm_range3}
\left|m-1+\cos k_y+\sqrt{1+[\sin k_y \sigma^y+b_y(z)]^2}\right|<1.
\end{align}
For the bound state in $z\sim \pm z^0$, one can safely rewrite Eq.\ \eqref{sm_range3} as
\begin{align}
|m+\cos k_y|<1.
\label{sm_range4}
\end{align}
Note that, Eq.\ \eqref{sm_range4} is identical to the condition on the range of the surface state at $k_z=0$. For $-2<m<0$, this range is centered around $k_y=0$ and has the form $-\Lambda_m<k_y<\Lambda_m$.

\begin{figure}
\includegraphics[width=0.5\columnwidth]{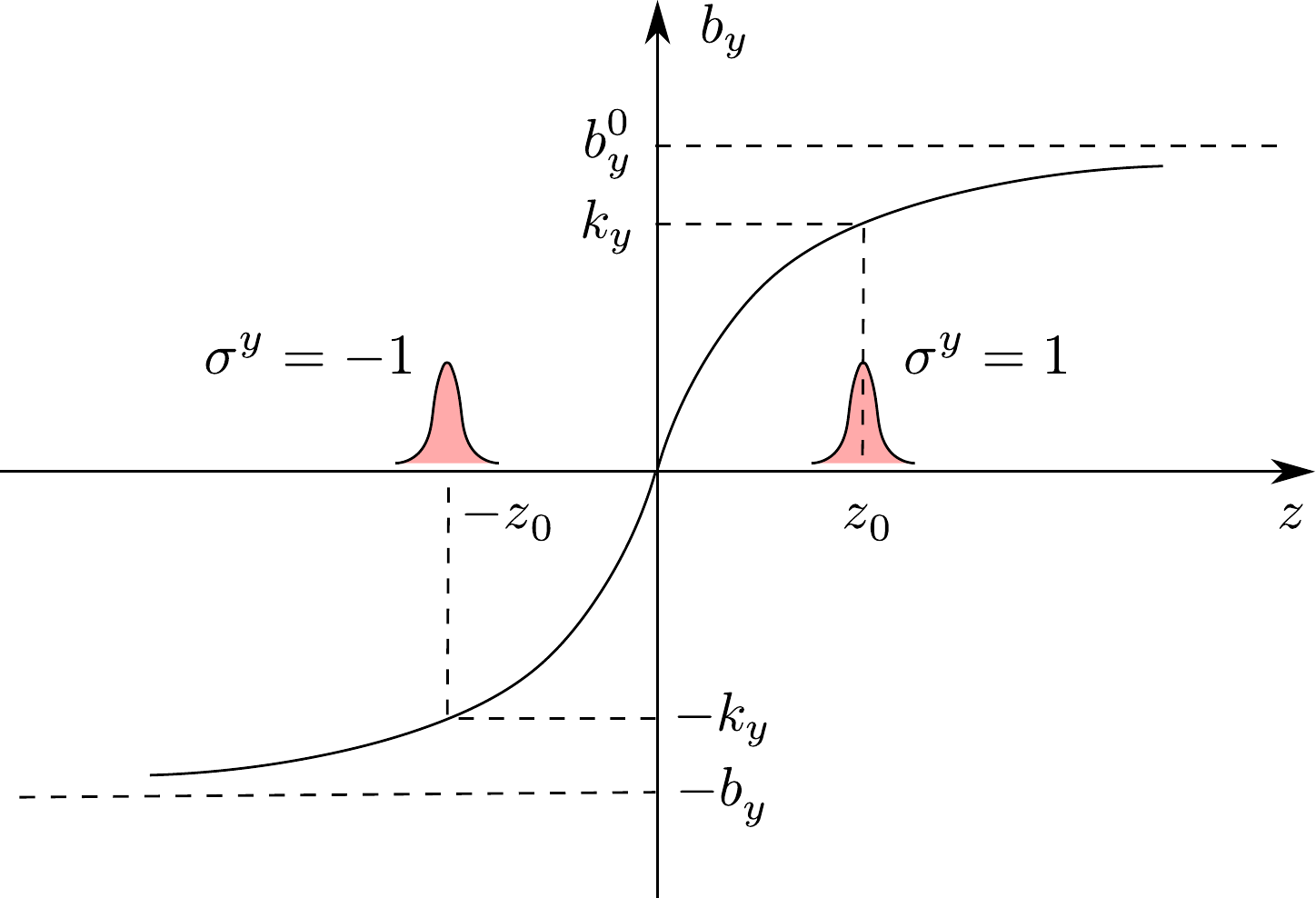}
\caption{The two zero-energy bound states at a given momentum $k_y$ localized at the domain wall in $b_y$ along $z$ direction, each corresponding to $\sigma^y=\pm1$.}
\label{sm_bst}
\end{figure}

\subsection{C.~~~Charge density from the domain wall bound states }

In the continuum limit, we can linearize Eq.\ \eqref{sm_fz} and neglect the upper cutoff on $k_y$. We take $f(z)=k_y+b_y(z)\sigma^y$, and from the ansatz \eqref{sm_asz} one can easily make sure this corresponds to two bound states localized at $z=\pm z_0$ for $\sigma^y=\pm1$, with $z_0$ given by $b_y(z_0)=k_y$. We illustrate this in Fig.\ \ref{sm_bst} for a given $k_y$. Each bound state contributes a charge $-e/2\sgn(m_A)$. On the other hand, for a domain wall across which $b_y$ changes from $-|b_y^0|$ to $|b_y^0|$, the total number of such bound states is $2\times |b_y^0|L_y/\pi$, where $L_y$ is the system size in the $y$-direction. Therefore, the total charge density bound at the domain wall of $b_y$ around $z=0$ is $-e|b_y^0|\sgn(m_A)/\pi$. This agrees with the result from the analytical response.

\begin{figure}[h]
\includegraphics[width=0.5\columnwidth]{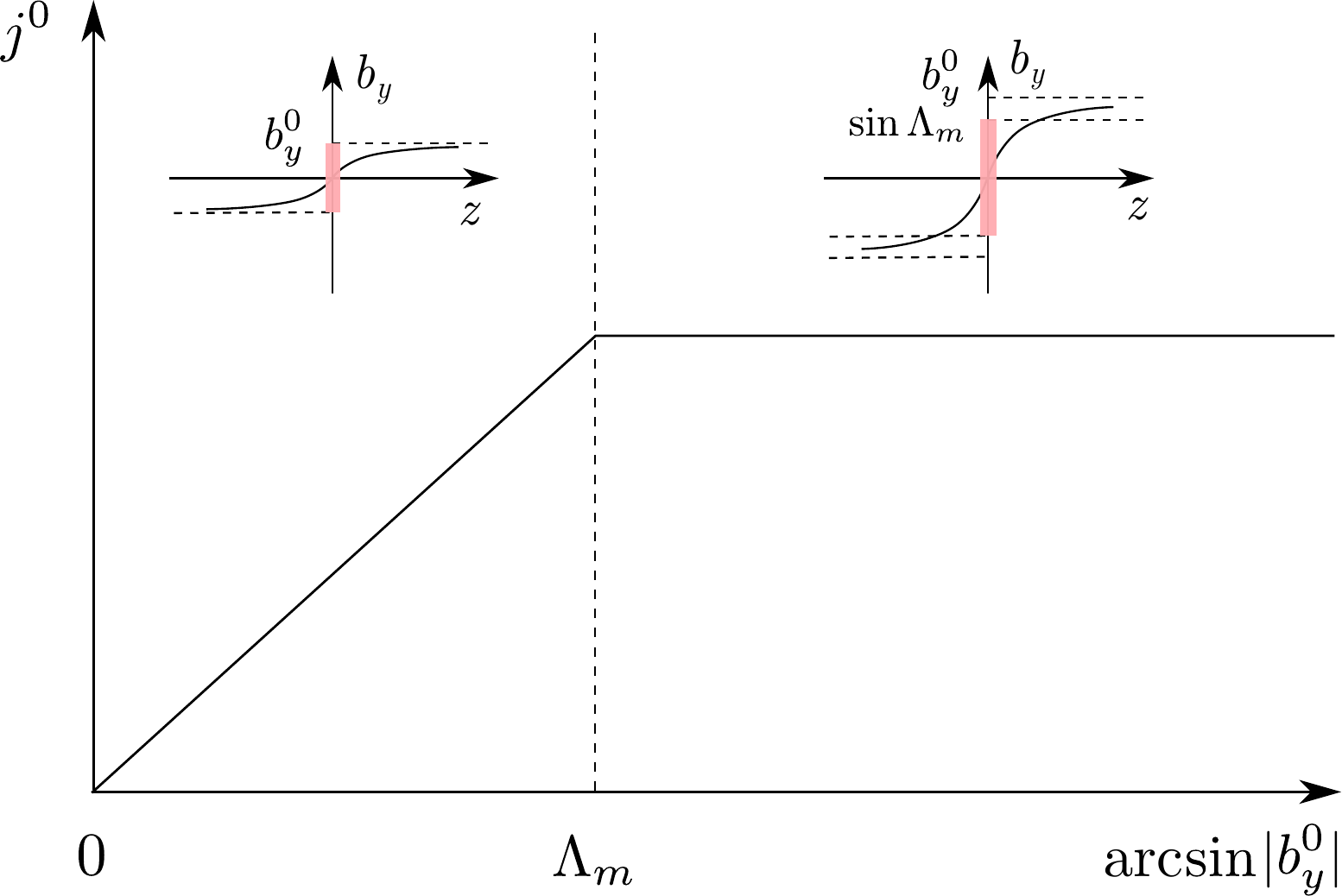}
\caption{Main figure: the charge density response $j^0$ at the surface domain wall of $b_y$ as a function of the height of the domain wall. In the first region the response is consistent with the analytical result obtained from the effective action. Insets: The domain walls of $b_y$ in $z$ direction. The colored thick lines mark the intermediate $b_y(z_1)$ values that have bound states localized at $z=\pm z_1$. }
\label{sm_resp}
\end{figure}

On the other hand, for a domain wall of $b_y$ with a larger ``height", whether the ansatz \eqref{sm_asz} correponds to a bound state on the surface is more tricky. ``Inside" the domain wall, $b_y=b_y(z)$ smoothly extrapolates from $-|b_y^0|$ to $|b_y^0|$. At an intermediate position $z=z_1$, if $b_y(z_1)$ is sufficiently large, there can be two possible situations. Within the range $-\Lambda_m<k_y<\Lambda_{m}$, the equation $f(z_1)=\sin k_y +b_y(z_1)\sigma^y=0$ can either have no solution, or have two solutions at which $f(z_1)$ has opposite slopes. In both cases, the ansatz \eqref{sm_asz} is not normalizable in $z$ direction, and hence does not correspond to a bound state solution localized at this $z=z_1$. 


Note that the two cases are closely related to the fate of the surface Dirac nodes discussed in Section II. Indeed, it is not difficult to see that the first case corresponds to when the surface Dirac nodes are eliminated within the range given by Eq.\ \eqref{sm_range}, and the second case corresponds to when additional Dirac nodes with opposite helicities to the original ones are introduced. In both cases, the topology of the bulk Hamiltonian changes. Therefore, we conclude that inside the domain wall, if an intermediate value $b_y(z_1)$ removes or cancels the surface Dirac nodes, there is {\it no} charge density bound at $z=z_1$. This occurs for 
\begin{align}
|b_y(z_1)|>|b_y^{cr}|=\sin \Lambda_m,
\end{align}
and we remember that for $-2<m<0$, $\sin \Lambda_m=\sqrt{-m(m+2)}$.

We can now obtain the generic charge density response as a function of the height of the domain wall $|b_y^0|$, which we plot in Fig.\ \ref{sm_resp}. For a small $|b^0_y|<|b_y^{cr}|=\sin \Lambda_m$, everywhere inside the domain wall there exist surface bound states, with $k_y$ ranging from $-\arcsin|b_y^0|$ to $\arcsin|b_y^0|$,
 leading to a charge density proportional to the corresponding momentum range $\arcsin|b_y^0|$, i.e. 
\be 
j^0=-\frac{e}{\pi}\arcsin|b_y^0|\sgn(m_A)\equiv-\frac{e}{\pi}|\tilde b_y^0|\sgn(m_A).
\ee
This response is universal, as it depends on universal numbers and a purely geometrical quantity $\tilde b_y^0\equiv\arcsin|b_y^0|$, which is the magnitude of the \emph{shift} of surface Dirac nodes on both sides of the domain wall. For $|b^0_y|>|b_y^{cr}|=\sin \Lambda_m$, however, only part of the domain wall interior traps bound states (see the second inset of Fig.\ \ref{sm_resp}). The charge response is $j^0=-e\sgn(m_A)/\pi\times\sin \Lambda_m$ and does not depend on $b_y^0$.

For a generic case given by a different $m$, or even a totally different lattice model, the value of $\sin\Lambda_m$ differs from our result. However, the behavior of the charge response remains the same, namely, the charge density first scales linearly with the height of the domain wall and then saturates at a critical value of $|b_y^0|$. This critical value of $|b_y^0|$ precisely corresponds to a change in the bulk topology, which leads to the elimination or cancellation of the surface Dirac nodes.

\end{document}